\documentclass[fleqn,usenatbib]{mnras}
\usepackage{graphicx}
\usepackage{natbib}
\usepackage{txfonts}
\usepackage{lastpage}
\usepackage{graphics,wrapfig,times}
\usepackage[absolute]{textpos}
\usepackage{tikz}
\usetikzlibrary{shapes.misc, positioning}
\usepackage[absolute]{textpos}
\usetikzlibrary{calc}

\begin{document}

\title[Three new 2+2 with inclination changes]{Three new 2+2 quadruple systems with changing inclination}

 \author[Zasche et al.]{Zasche,~P.$^1$ \thanks{E-mail: zasche@sirrah.troja.mff.cuni.cz},
 Henzl,~Z.$^{2,3}$,
 Ma\v{s}ek~M.$^{3,4}$,
 K\'{a}ra~J.$^{1,7}$,
 Ku\v{c}\'{a}kov\'{a}$^{1,3,5,9}$,
 Merc~J.$^{1,8}$,
 Uhla\v{r},~R.$^{6}$\\
 $^1$ Charles University, Faculty of Mathematics and Physics, Astronomical Institute, V Hole\v{s}ovi\v{c}k\'ach 2, Praha 8, \\ \hspace{0.5cm} 180 00, Czech Republic \\
 $^{2}$ Hv\v{e}zd\'arna Jaroslava Trnky ve Slan\'em, Nosa\v{c}ick\'a 1713, Slan\'y 1, 274 01, Czech Republic \\
 $^{3}$ Variable Star and Exoplanet Section, Czech Astronomical Society, Fri\v{c}ova 298, 251 65 Ond\v{r}ejov, Czech Republic \\
 $^{4}$ FZU - Institute of Physics of the Czech Academy of Sciences, Na Slovance 1999/2, CZ-182~21, Praha, Czech Republic\\
 $^{5}$ Research Centre for Theoretical Physics and Astrophysics, Institute of Physics, Silesian University in Opava, Bezru\v{c}ovo n\'am. 13, CZ-746 01, \\ \hspace{0.5cm} Opava, Czech Republic\\
 $^{6}$ Private Observatory, Poho\v{r}\'{\i} 71, CZ-254 01 J\'{\i}lov\'e u Prahy, Czech Republic \\
 $^{7}$ Department of Physics and Astronomy, University of Texas Rio Grande Valley, Brownsville, TX 78520, USA \\
 $^{8}$ Instituto de Astrofisica de Canarias, Calle Via Lactea, s / n, E-38205 La Laguna, Tenerife, Spain \\
 $^{9}$ Astronomical Institute, Academy of Sciences, Fri\v{c}ova 298, 251 65, Ond\v{r}ejov, Czech Republic \\
 }


\date{\today}
\pagerange{\pageref{firstpage}--\pageref{LastPage}} \pubyear{2025} \maketitle \label{firstpage}

\begin{abstract}
We present a unique discovery of three new detected systems showing two different phenomena
together. These are 2+2 quadruple stellar systems showing two eclipsing binaries as the inner
pairs. And besides that, these systems were also found to exhibit the precession of the inner
orbits causing the inclination changes manifesting themselves through the eclipse depth
variations. We are not aware of any similar known system on the sky nowadays, hence our discovery
is really unique. In particular these systems are: CzeV4315 = HD 228777 (periods 6.7391 d and
0.91932 d, inclination change of pair B of about 1.4$^{\circ}$/yr); ASASSN-V J075203.23-323102.7 =
GDS$\_$J0752031-323102 (8.86916 d + 2.6817 d, inclination change of pair B of about
1.03$^\circ$/yr, now only ellipsoidal variations);  ASASSN-V J105824.33-611347.6 = TIC 465899856
(2.3304 d + 13.0033 d, inclination change of pair B, now undetectable). These systems provide us
unique insight into the quadruple-star dynamics, including the orbit-orbit interaction,
Kozai-Lidov cycles, and testing the stellar formation theories of these higher-order multiples.
\end{abstract}



\section{Introduction}

Sometimes, the eclipsing binaries are presented as standstones of modern stellar astrophysics due
to their role in the 20-th century. They are being used to precisely measure the stellar radii and
masses, as well as to calibrate the distance ladder, and the stellar evolution theories via
looking into the stellar interiors. About all of these aspects many different papers were
published so far \citep{2012ocpd.conf...51S,2010A&A...519A..57C}.

However, nowadays in the era of large surveys and ultra-precise satellite data it is not even
possible to go into thorough in-deep analysis of every single eclipsing binary system (hereafter
EB) since their number is now over 2 million \citep{2023A&A...674A..16M}. Hence, it would be more
useful to focus in more detail into some specific types of EBs, which are limited in number, and
are of high astrophysical importance, for example offer us some unique insight into processes or
effects otherwise hardly visible. It means: objects observable, but rare enough, showing different
and unique astrophysical phenomena.

We think that the multiply-eclipsing systems are definitely such type of objects, especially when
they also present a rare precession of the orbits of their inner binaries. Among the known more
than 2 million EBs there are only about 1000 known candidates for doubly eclipsing systems. I.e.
those systems, where one point source on the sky produces two different eclipsing periods.

On the other hand, there are still only a handful of EBs showing orbital precession. With the
precession, we mean the slow changes of inclination (hence the eclipse depth variations) caused by
the third component in the system. In the review of the topic of multiples among the EBs published
by \cite{2022Galax..10....9B} were listed only 34 such stars known in our Galaxy, and 58
candidates were found in LMC \& SMC galaxies from the OGLE surveys by \cite{2018A&A...609A..46J}.
Since then, eight new candidates from the TESS satellite \cite{2015JATIS...1a4003R} were published
\citep{2024A&A...685A..43M}. It means the number of such inclination changing EBs is still only
about 100 on the whole sky and are very rare among the stellar systems. And finally, there are
even fewer such ones where the third body causing the orbital precession was detected and the
orbit was derived: only 25 up to date.

Nevertheless, if we take a look at the overlap of these two groups of multiple stellar systems, we
find zero. It means up to now there was not a single detection of precession EB in the 2+2
quadruple doubly eclipsing system. In our paper, we present three such unique examples.

\section{Selection of targets}

All of the targets were detected as a by-product of our long-term effort of identifying new
quadruples of doubly eclipsing type. It means that during our scanning of the TESS data in new
sectors we found several interesting systems showing two different eclipsing periods and the shape
of the two eclipses changes significantly over the time interval of available data.

\begin{table*}
  \caption{Basic information about the systems.}  \label{systemsInfo}
  \begin{tabular}{c c c c c c c}\\[-3mm]
\hline \hline\\[-3mm]
  Target name                         &  Other name            &      TESS      &     RA     &     DE     & Mag$_{max}$ $^{\star}$ & Temperature/sp.type        \\
                                      &                        & identification &  [J2000.0] & [J2000.0]  &                        & information $^{\star\star}$ \\
 \hline
 CzeV4315                    & HD 228777              & TIC 12248375   & 20 17 40.3 & +39 14 35.2 &  9.56 (V) & sp B9$^{\sharp}$   \\
 ASASSN-V J075203.23-323102.7 & GDS$\_$J0752031-323102 & TIC 150789352  & 07 52 03.1 & -32 31 02.2 & 12.52 (V) & $T_{eff} = 5739$ K  \\
 ASASSN-V J105824.33-611347.6 & UCAC4 144-063891       & TIC 465899856  & 10 58 23.9 & -61 13 46.0 & 12.67 (V) & sp O6.5IV((f))$^{\sharp\sharp}$  \\
  \hline
\end{tabular}
 {\small Notes: $^\star$ Out-of-eclipse magnitude, 
 $^{\star\star}$ Effective temperature taken from the Gaia DR3 catalogue \citep{2023A&A...674A...1G},
 $^{\sharp}$ \cite{1961AbaOB..26...35K},  $^{\sharp\sharp}$ \cite{2022A&A...657A.131M}.
 }
\end{table*}

No special focus on early or late type stars was taken since we deal with only photometry and no
detailed spectroscopic study is available for the presented systems. Typically, the systems have
one prominent period, which is deep enough and was sometimes even detected earlier. See below for
details on particular systems. However, the secondary period detection is novel here and this
additional eclipses were the ones showing orbital precession demonstrated via eclipse depth
variations.

The fact that the inclination is changing is being explained in a standard way as the third-body
interaction of additional body in the system on non-coplanar orbit with the inner eclipsing binary
(see e.g. \citealt{1975A&A....42..229S}). Such a distant body also shows orbital precession, but
it is usually undetectable, since what we typically see is the inner eclipsing body and not the
third component itself. Changing depth of eclipses is only a consequence of this motion. In most
cases the inner eclipsing binary is the more massive dominant pair and the third body is usually
smaller. The first one such system was IU Aur, where the changes of inclination were detected by
\cite{1971BAICz..22..168M}. The most typical examples are systems studied in the past like V907
Sco \citep{2023AJ....165...81Z}, HS Hya \citep{2022AJ....163...94V}, or SS Lac
\citep{2001AJ....121.2227T}. For the inclination precession to be detectable, it should be rapid
enough. This means that the rate of inclination change should be of the order of a few degrees
over the time span of available (mostly TESS) data. Such a third body should therefore orbit with
relatively short period (due to the fact that the precession periodicity is proportional to
$P_{AB}^2/P_1$, where $P_{AB}$ is the outer period, while $P_1$ is the inner one), definitely
shorter than several years.

However, here we deal with quite a different situation that has never been mentioned before. Our
third distant component is not a single star, but also a binary showing eclipses. Hence this is
for the first time we can also say something about this body causing the precession itself. We can
model its eclipses (at least theoretically) to infer some basic physical or orbital parameters.
For almost all other known inclination-changing systems any information about the third component
is usually missing completely and its parameters are being only derived using some dynamical
modelling of the change of eclipses of pair A. The only one exception here was the discovery and
modelling of the triple KIC 7955301 \citep{2022A&A...668A.173G}, where the third component is a
dominant red giant star. Quite challenging (and definitely fruitful for any dynamical four-body
modelling of the system) would be the detection of changes of the eclipse depth also for the other
pair in the future.

The summary of some basic information about the individual systems is in Table \ref{systemsInfo},
where the stars are being sorted from the best-studied to the least studied one, and from the
brightest to the faintest star as well.

\section{Data handling}

The data from TESS were the primary source of information about the stars and also used for
modelling for both of the inner binaries. However, besides that we also tried to use another older
photometry to enlarge the time span of available data to better cover the inclination changes
(similarly as was done for HS Hya by \citealt{2021AJ....162..189D}) as well as also to obtain new
observations with our means. However, there was a problem that some of the stars are relatively
faint and their photometric amplitudes small. For most of the systems the changing pair B was the
shallower one, and these shallow eclipses can only hardly be identified on older less precise
survey photometry.

Sometimes the older photometry was used only for the confirmed detection of the more prominent
deeper eclipses and stability of its orbital period. However, we are aware of the fact that such a
complicated and dynamically interacting systems can also show very complex curvatures in the ETV
(eclipse timing variations) diagrams as was shown e.g. by \cite{2022MNRAS.515.3773B}. Hence, to
easily construct the phased light curve should not be so easy as for some more traditional
eclipsing binaries.

We are also aware of the fact that the TESS data in different sectors may in principal have
different level of light contamination, from the close-by sources. Hence, also the third light
fraction has to be taken as a free independent parameter in all of the TESS sectors independently.
The level of the resulting third light values as obtained from analysing both A and B binaries can
in principle give us some estimation of the light ratio of the two inner doubles and also about
the flux distribution among all four components in the system for some prospective future
spectroscopic detection of all the components in the system.

\section{Method for light curve analysis}

For the whole process of analysis of the light curves, we used a similar approach as for the
doubly eclipsing systems studied recently (see e.g. \citealt{2024A&A...687A...6Z},
\citealt{2020A&A...642A..63Z}).

This means that at first we chose the best sector of the TESS data. From this best sector the
light curve analysis was done. Now in our present study, the situation is a bit more complicated
due to changing eclipse depth. Hence, the selected TESS sector was chosen to be the one with the
best quality (i.e. typically with the highest cadence) as well as with the most pronounced
eclipses of both pairs, if possible.

After than, the first preliminary light curve fit of the pair A was performed, producing the
residuals with only the pair B variations. All of the light curve fittings were done with the
Wilson-Devinney based software {\sc PHOEBE} \citep{2005ApJ...628..426P}. The light curve fit of
pair B was done with some preliminary value of the detected period of pair B. The fit of B was
then used again for the complete photometry, from which it was subtracted to obtain only the pair
A with higher accuracy. Both pairs A and B were used also for the better estimation of their
orbital periods. Both periods were computed using the AFP method \citep{2014A&A...572A..71Z},
which derives the times of eclipses over longer time intervals. This procedure was repeated
several times to obtain a reasonable fits of both pairs - their light curves as well as their
precise periods.

Typically we started with the assumption of equal luminosities of both pairs (i.e. the third light
fraction of 50\%). After then, the third light was freed from constraints, as was also the mass
ratio (from starting value of 1.0) for some of the stars with better data and larger
out-of-eclipse variations.

 \section{Individual systems}

In this section, we summarize our findings about the three analysed systems. The first one will be
presented in more detail, the rest only briefly mentioned and only the most interesting findings
for them will be given.

\subsection{CzeV4315}

The first one and definitely the most interesting is the star named CzeV4315 in the Czech Variable
Star Catalogue \citep{2017OEJV..185....1S} due to its discovery as a variable by Z.Henzl in 2022.
It is also included in other catalogues under other names: HD~228777 = BD+38~3992 =
GSC~03151-00289. Despite its location in Cygnus constellation, it was not observed with the Kepler
satellite. The star was also resolved as a double and is included in the Washington Double Star
Catalogue (WDS), where the three listed astrometric observations show only very small or even
negligible movement. Hence, its period is probably very long, probably thousands of years. Due to
its relatively high brightness, the star was also classified spectroscopically. However, different
publications give slightly different spectral types, ranging from A0 (oldest publication by
\citealt{1925AnHar.100...17C}) to B9 (newest by \citealt{1961AbaOB..26...35K}).

 \begin{figure}
  \centering
     \begin{picture}(380,300)
        \put(10,-12){ \includegraphics[width=0.41\textwidth]{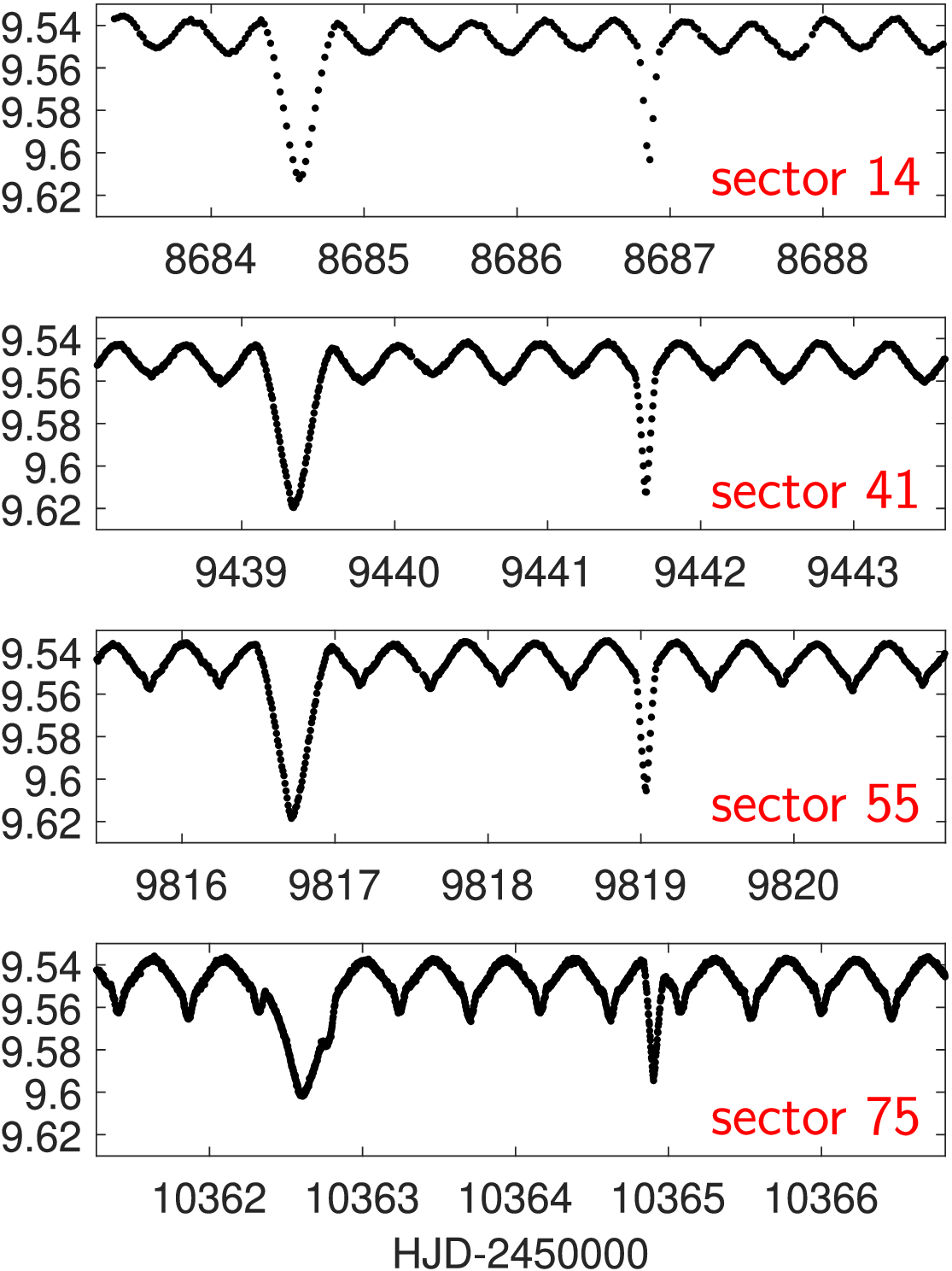}}
        \put(-10,120){ {\large \textsf{\rotatebox{90}{TESS Magnitude}}}}
     \end{picture}
  \caption{Star CzeV4315: Snapshot of TESS data over various sectors (from top to bottom: sectors 14, 41, 55, and 75). Increasing depth of the eclipses of pair B is clearly visible here, making the combined shape of A+B more complicated.}
  \label{FigLC_czev4315_sectors}
 \end{figure}

There exist several TESS sectors of data, covering a time interval of 5 years. This is a long
enough time base for a rapid inclination precession to emerge. For the first two TESS sectors
there appears that the star is an eclipsing binary of about 6.74 days period. Additional small
variations on the light curve (hereafter LC) can easily be attributed to some pulsations of either
of the components.

However, after the next release of additional sectors of TESS data, the picture of the whole
system changed completely. From this point we call the dominant 6.7-d variation "pair A". Such an
eclipsing-shape variation remained almost unchanged. However, the shallow additional variation
became more obvious also eclipsing (we call this "pair B") with the periodicity of about 0.919
days. The amplitude of this pair B is increasing over the whole interval of the TESS data. Now the
eclipses are clearly visible. All this behaviour is presented in our Figure
\ref{FigLC_czev4315_sectors}.

\begin{figure}
 \centering
 \begin{picture}(380,330)
 \put(14,229){
  \includegraphics[width=0.38\textwidth]{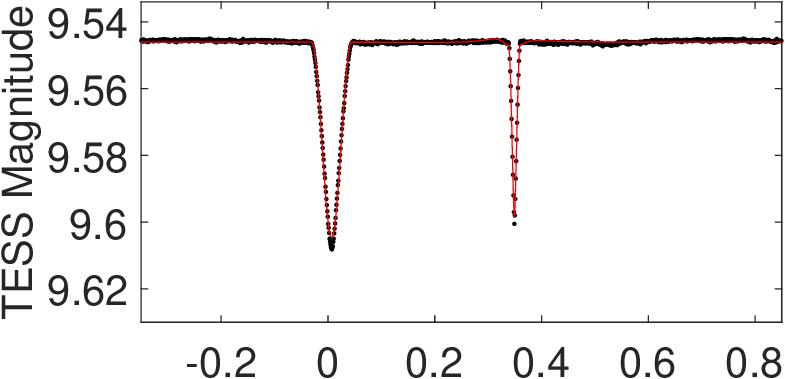}}
  \put(10,0){
  \includegraphics[width=0.39\textwidth]{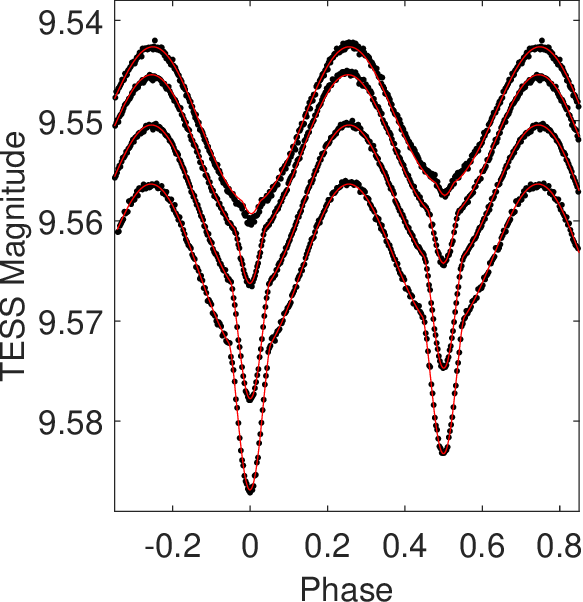}}
   \put(155,248){ {\Large \textsf{Pair A}}}
   \put(155,40){ {\Large \textsf{Pair B}}}
  \end{picture}
  \caption{Light-curve fits of both pairs of system CzeV4315. In the bottom plot are given the individual light curves from three different TESS sectors (from top to bottom: 41, 55, 75, and 81). Changing depth of eclipses is clearly visible here.}
  \label{FigLC_Czev4315}
 \end{figure}

In Figure \ref{FigLC_Czev4315} we present the disentangled shapes of both A and B binaries. The
dominant pair A is clearly a detached eclipsing binary orbiting on very eccentric orbit
(e$\approx$0.64). Such a high value of eccentricity is one of the highest among the binaries with
a similar orbital period of about 6.7 days. Whether this value was somehow excited to its current
value via some dynamical interactions in the multiple system we leave here as an open question. On
the other hand, the pair B is much more compact, but still a detached one. It has a circular orbit
and started to be eclipsing only a few years ago. Its increasing depths of eclipses are visible
even for a naked eye in Figure \ref{FigLC_Czev4315}.

With some easy assumptions (linear ephemerides over one TESS sector, constant shape of the light
curves, subtraction of particular light curve of one pair to obtain the LC of another pair) we
tried to do some preliminary LC analysis. The results of this analysis are summarized below in
Table \ref{LCparamAll}. We emphasize here once again that these are still rather preliminary
results. The most interesting is definitely the evolution of inclination value of the B pair.
According to this rapid movement of the orbital plane and our inferred luminosity ratios of both
pairs, there seems to be one unevitable result: the pair B will be the dominant one in a few years
from now (it means producing deeper eclipses than pair A in the combined light curve).

\begin{figure}
  \centering
   \includegraphics[width=0.38\textwidth]{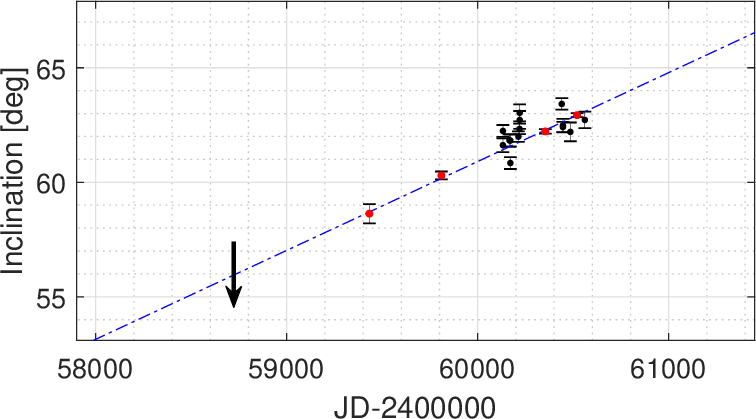}
  \caption{Star CzeV4315: inclination change of pair B. The red dots are from TESS data, black ones from our ground-based photometry. Blue dash-doted curve is only the simple linear fit.}
  \label{czev4315_incl}
 \end{figure}

The time evolution of the inclination values of pair B is also plotted in Figure
\ref{czev4315_incl}. As one can see, the TESS data are of much higher accuracy, but the overall
trend of the inclination change of pair B is pretty visible now also on our ground-based
observations. These were obtained on three different sites: Ond\v{r}ejov Observatory in Czech
Republic with its 65-cm telescope; small 30-cm telescope at the private observatory in
Velt\v{e}\v{z}e u Loun, Czech Republic (Z.H.), and with the 15-cm telescope at the private
observatory (by R.U.) in J\'{\i}lov\'e u Prahy, Czech Republic. Unfortunately, due to missing
eclipses on the first TESS sectors only some inclination limit can be derived (this is being
indicated with an arrow sign in Fig.\ref{czev4315_incl}).

\begin{table*}
  \caption{The light curve parameters for both A and B pairs of the analysed systems.}  \label{LCparamAll}
  \centering
  \scalebox{0.95}{
\begin{tabular}{c @{\vline} c  c @{\vline} c  c @{\vline} c  c }
   \hline\hline
  System & \multicolumn{2}{c@{\vline}} {CzeV4315}                               & \multicolumn{2}{c@{\vline}}{ASASSN-V J075203.23-323102.7} & \multicolumn{2}{c}{ASASSN-V J105824.33-611347.6}    \\ 
  \multicolumn{1}{c@{\vline}}{ } & {\sc p a i r \,\, A} & {\sc p a i r \,\, B}  &  {\sc p a i r \,\, A}       & {\sc p a i r \,\, B}        &  {\sc p a i r \,\, A}   & {\sc p a i r \,\, B}      \\ \hline
 $P$ [d]                & 6.73907 $\pm$ 0.00025    & 0.9193238 $\pm$ 0.0000015\,& \,8.8691638 $\pm$ 0.0000846 & 2.6817062 $\pm$ 0.0000263\, & \,2.3304383 $\pm$ 0.000061& 13.003275 $\pm$ 0.000308  \\
 $HJD_0$ [d]            & \,2459438.530 $\pm$ 0.013& 2460132.034 $\pm$ 0.001    & 2458508.751 $\pm$ 0.082     & 2458506.299 $\pm$ 0.002     & 2458809.252 $\pm$ 0.032 & 2459314.683 $\pm$ 0.105   \\
 $e$                    & 0.64 $\pm$ 0.04          & 0.0                        &  0.21 $\pm$ 0.04            &  0.069 $\pm$ 0.009          & 0.051 $\pm$ 0.010       & 0.24 $\pm$ 0.02           \\
 $\omega$ [deg]         & 252.2 $\pm$ 2.6          & --                         &  91.3 $\pm$ 1.1             &  70.5 $\pm$ 5.8             & 235.1 $\pm$ 8.2         & 56.69 $\pm$ 2.45          \\
 $i$ [deg]              & 87.44 $\pm$ 0.21         & var.                       &  80.74 $\pm$ 0.09           &  var.                       & 76.48 $\pm$ 0.75        & var.                      \\
 $q=\frac{M_2}{M_1}$    & 0.87 $\pm$ 0.06          & 1.02 $\pm$ 0.02            &   1.0 (fixed)               &   1.0 (fixed)               & 0.95 $\pm$ 0.04         &  1.0 (fixed)              \\
 $T_1$ [K]              & 10000 (fixed)            & 10000 (fixed)              &   6000 (fixed)              &   6000 (fixed)              & 40000 (fixed)           &  40000 (fixed)            \\
 $T_2$ [K]              & 7236 $\pm$ 65            &  9735 $\pm$ 48             &   6101 $\pm$ 54             &   6195 $\pm$ 72             & 30623 $\pm$ 328         & 32481 $\pm$ 507           \\
 $R_1/a$                & 0.070 $\pm$ 0.003        & 0.275 $\pm$ 0.004          &   0.128 $\pm$ 0.003         &   0.136 $\pm$ 0.004         & 0.223 $\pm$ 0.005       & 0.123 $\pm$ 0.004         \\
 $R_2/a$                & 0.066 $\pm$ 0.003        & 0.273 $\pm$ 0.004          &   0.137 $\pm$ 0.003         &   0.182 $\pm$ 0.004         & 0.183 $\pm$ 0.004       & 0.076 $\pm$ 0.004         \\
 $L_1$ [\%]             & 13.6 $\pm$ 0.8           & 19.2 $\pm$ 0.4             &   28.5 $\pm$ 0.4            &    5.3 $\pm$ 0.7            & 6.5 $\pm$ 1.0           & 22.8 $\pm$ 2.7            \\
 $L_2$ [\%]             &  4.9 $\pm$ 0.5           & 17.7 $\pm$ 0.3             &   40.5 $\pm$ 0.6            &   10.7 $\pm$ 0.8            & 2.5 $\pm$ 0.7           &  5.5 $\pm$ 0.6            \\
 $L_3$ [\%]             & 81.7 $\pm$ 1.4           & 63.1 $\pm$ 0.7             &   31.0 $\pm$ 0.9            &   84.0 $\pm$ 2.1            & 91.0 $\pm$ 2.2          & 71.7 $\pm$ 4.0            \\ \hline\hline
 \end{tabular}}\\
\end{table*}

\begin{table}
  \caption{The inclination angles for CzeV4315$\_$B.}  \label{Tab_CzeV4315_incl}
  \centering 
\begin{tabular}{c | c | c }
   \hline\hline 
  \multicolumn{3}{c}{\sc pair B of CzeV4315}        \\ \hline
 $HJD$ [mid] &  $i$ [deg]  & Source  \\ \hline
  24560520   &  62.93 $\pm$ 0.09 & TESS - sector 81 \\
  24560354   &  62.22 $\pm$ 0.10 & TESS - sector 75 \\
  24559810   &  60.30 $\pm$ 0.17 & TESS - sector 55 \\
  24559433   &  58.63 $\pm$ 0.42 & TESS - sector 41 \\
  24558725   &  $<$ 57.4         & TESS - sector 15 \\
  24560132   &  62.24 $\pm$ 0.26 & Z.H. \\
  24560132   &  61.62 $\pm$ 0.30 & Z.H. \\
  24560171   &  60.84 $\pm$ 0.26 & Ond\v{r}ejov \\
  24560171   &  61.82 $\pm$ 0.26 & Ond\v{r}ejov \\
  24560213   &  61.99 $\pm$ 0.23 & Ond\v{r}ejov \\
  24560219   &  62.33 $\pm$ 0.23 & Ond\v{r}ejov \\
  24560220   &  63.03 $\pm$ 0.37 & Ond\v{r}ejov \\
  24560440   &  63.42 $\pm$ 0.25 & Ond\v{r}ejov \\
  24560446   &  62.48 $\pm$ 0.29 & Ond\v{r}ejov \\
  24560446   &  62.41 $\pm$ 0.23 & Z.H. \\
  24560166   &  61.83 $\pm$ 0.27 & R.U. \\
  24560220   &  62.72 $\pm$ 0.39 & R.U. \\
  24560485   &  62.20 $\pm$ 0.41 & R.U. \\
  24560561   &  62.72 $\pm$ 0.36 & R.U. \\
 \hline
  \noalign{\smallskip}\hline
 \end{tabular}
\end{table}

\begin{figure*}
 \centering
 \begin{picture}(480,630)
    \put(1,577){ {\Huge \textsf{A:}}}
    \put(1,475){ {\Huge \textsf{B:}}}
   \put(30,532){\includegraphics[width=0.33\textwidth]{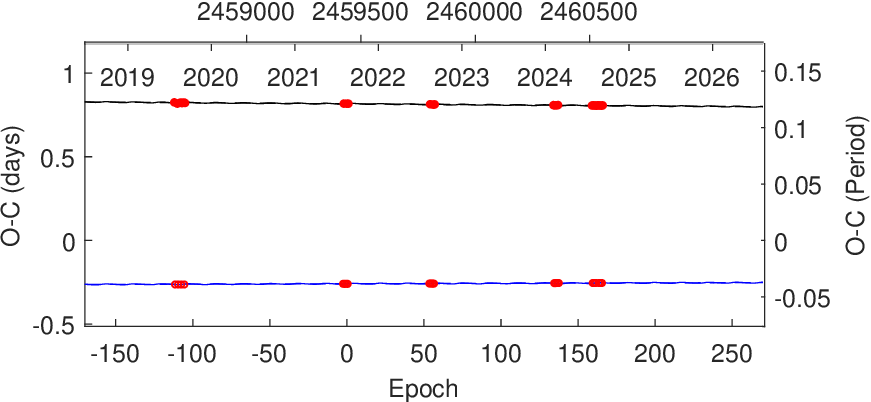}}
   \put(250,532){\includegraphics[width=0.33\textwidth]{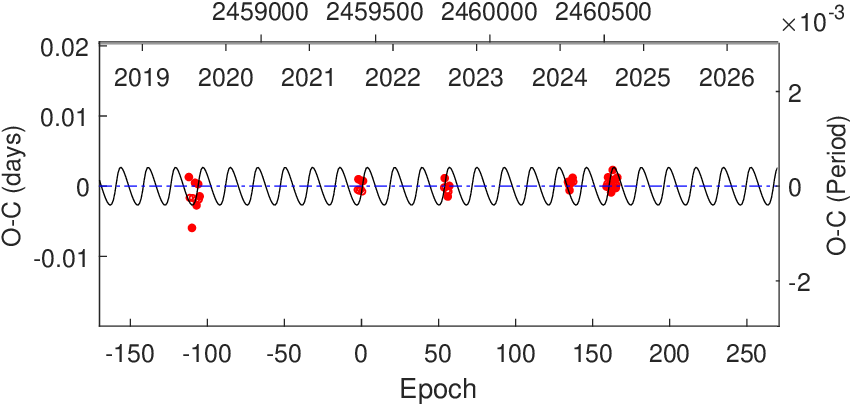}}
    \linethickness{0.7mm}  \put(1,426){\line(1,0){470}}
   \put(30,432){\includegraphics[width=0.33\textwidth]{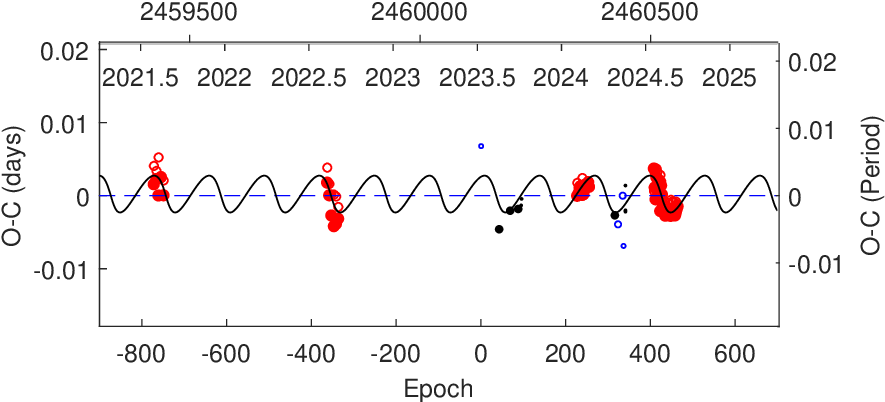}}
     \put(125,566){ {\Large \textsf{Pair A}}}
     \put(325,560){ {\Large \textsf{Pair A}}}
     \put(125,457){ {\Large \textsf{Pair B}}}
     \put(215,522){\begin{tikzpicture} \node (2) [draw, rounded rectangle] {\LARGE CzeV4315};\end{tikzpicture}}
    \put(1,365){ {\Huge \textsf{A:}}}
    \put(1,263){ {\Huge \textsf{B:}}}
   \put(30,325){\includegraphics[width=0.33\textwidth]{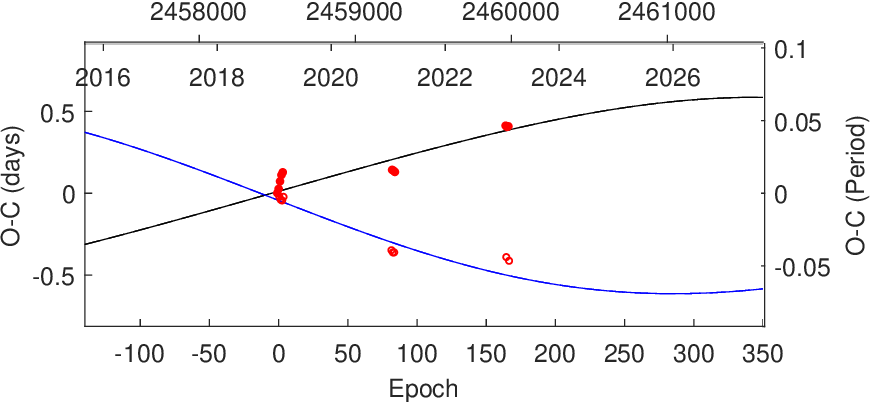}}
   \put(250,325){\includegraphics[width=0.33\textwidth]{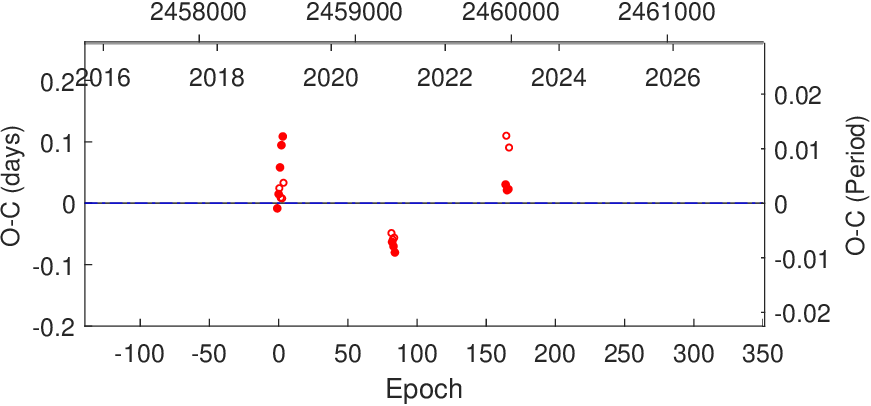}}
    \linethickness{0.7mm}  \put(1,214){\line(1,0){470}}
   \put(30,219){\includegraphics[width=0.33\textwidth]{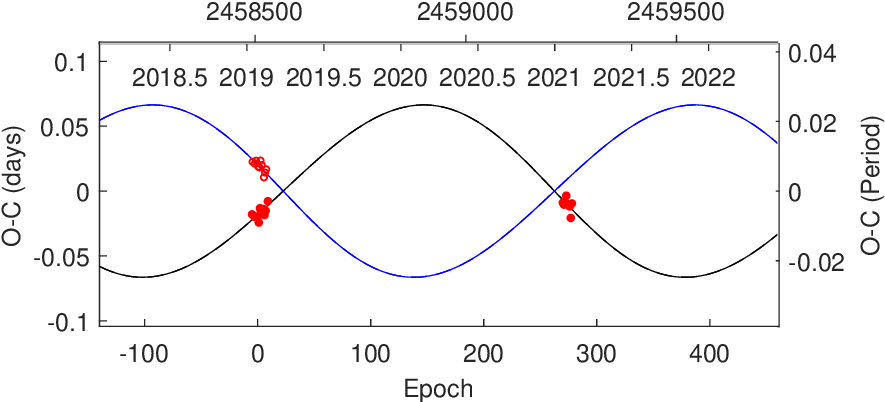}}
   \put(250,219){\includegraphics[width=0.33\textwidth]{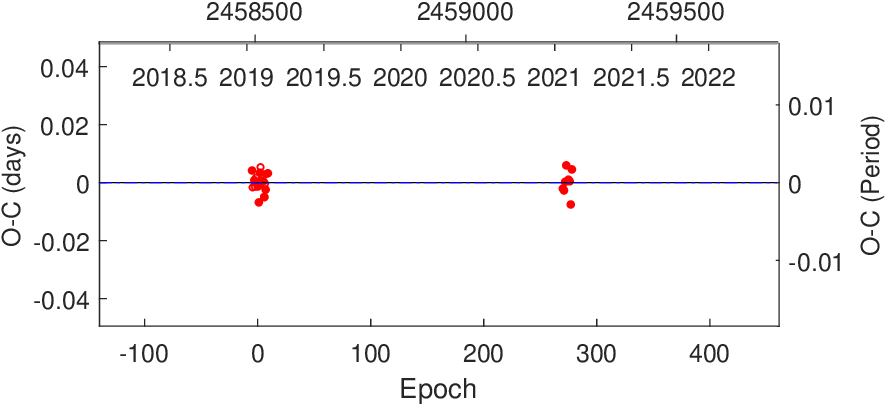}}
     \put(165,356){ {\Large \textsf{Pair A}}}
     \put(380,350){ {\Large \textsf{Pair A}}}
     \put(139,242){ {\Large \textsf{Pair B}}}
     \put(380,245){ {\Large \textsf{Pair B}}}
     \put(155,314){\begin{tikzpicture} \node (3) [draw, rounded rectangle] {\LARGE ASASSN-V J075203.23-323102.7};\end{tikzpicture}}
    \put(1,152){ {\Huge \textsf{A:}}}
    \put(1,050){ {\Huge \textsf{B:}}}
   \put(30,112){\includegraphics[width=0.33\textwidth]{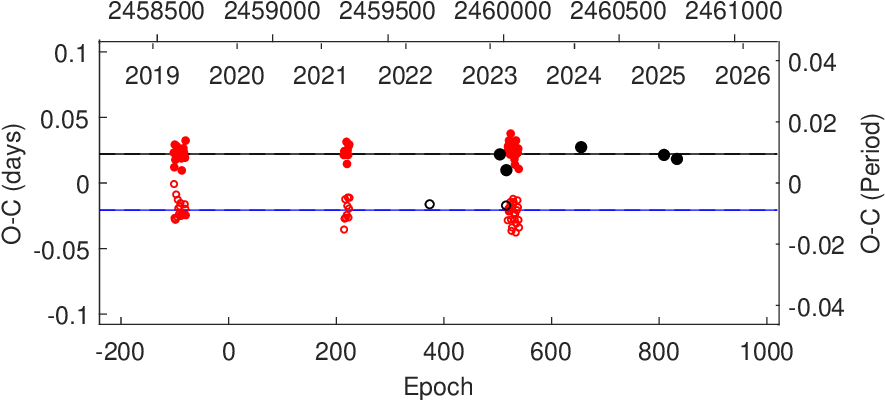}}
   \put(250,112){\includegraphics[width=0.33\textwidth]{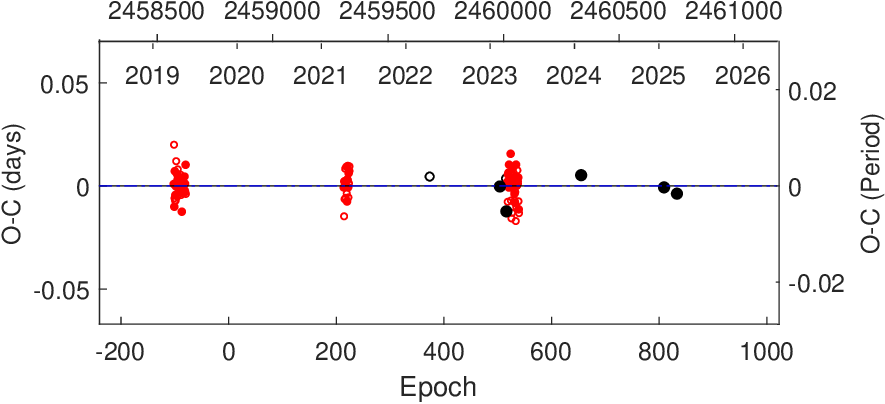}}
   \put(30,006){\includegraphics[width=0.33\textwidth]{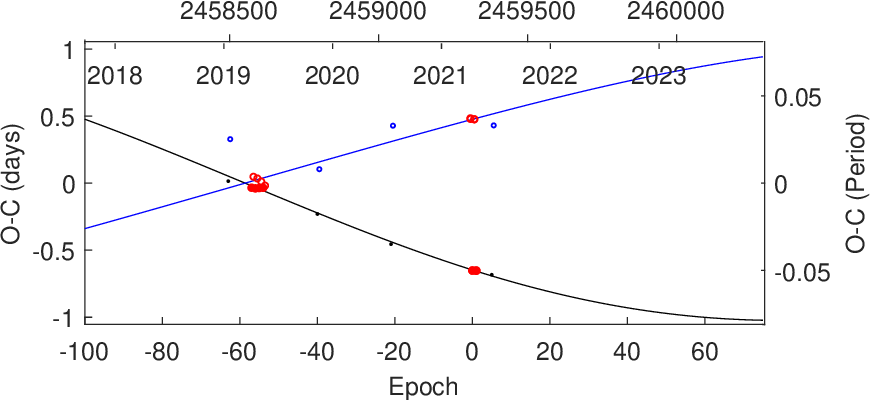}}
   \put(250,006){\includegraphics[width=0.33\textwidth]{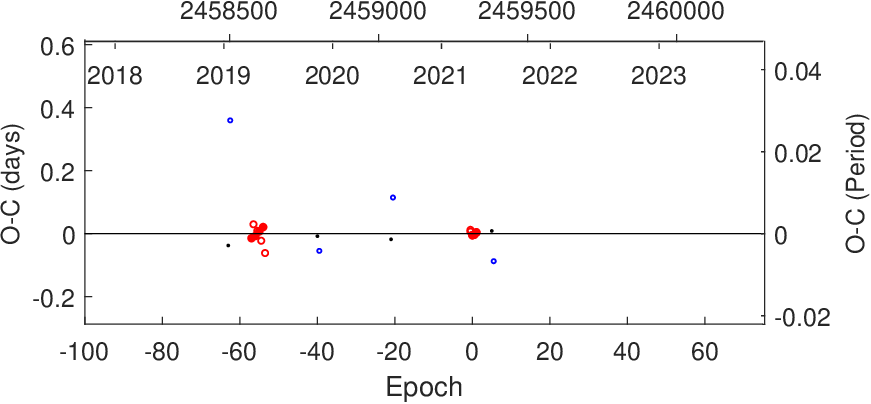}}
     \put(150,135){ {\Large \textsf{Pair A}}}
     \put(370,135){ {\Large \textsf{Pair A}}}
     \put(165,038){ {\Large \textsf{Pair B}}}
     \put(380,032){ {\Large \textsf{Pair B}}}
     \put(155,103){\begin{tikzpicture} \node (3) [draw, rounded rectangle] {\LARGE ASASSN-V J105824.33-611347.6};\end{tikzpicture}}
  \end{picture}
  \caption{ETV diagrams of both pairs for all of the studied systems. For each of them the upper plots are for the pair A, while bottom ones for pair B. Left-hand-side figures are the complete ETV diagrams, while the right-hand side are after subtraction of the apsidal motion. The red symbols (full dots represent primary eclipses, open circles the secondary ones) are the TESS data, the rest are from other surveys and our new data.}
  \label{Figs_ETV}
 \end{figure*}
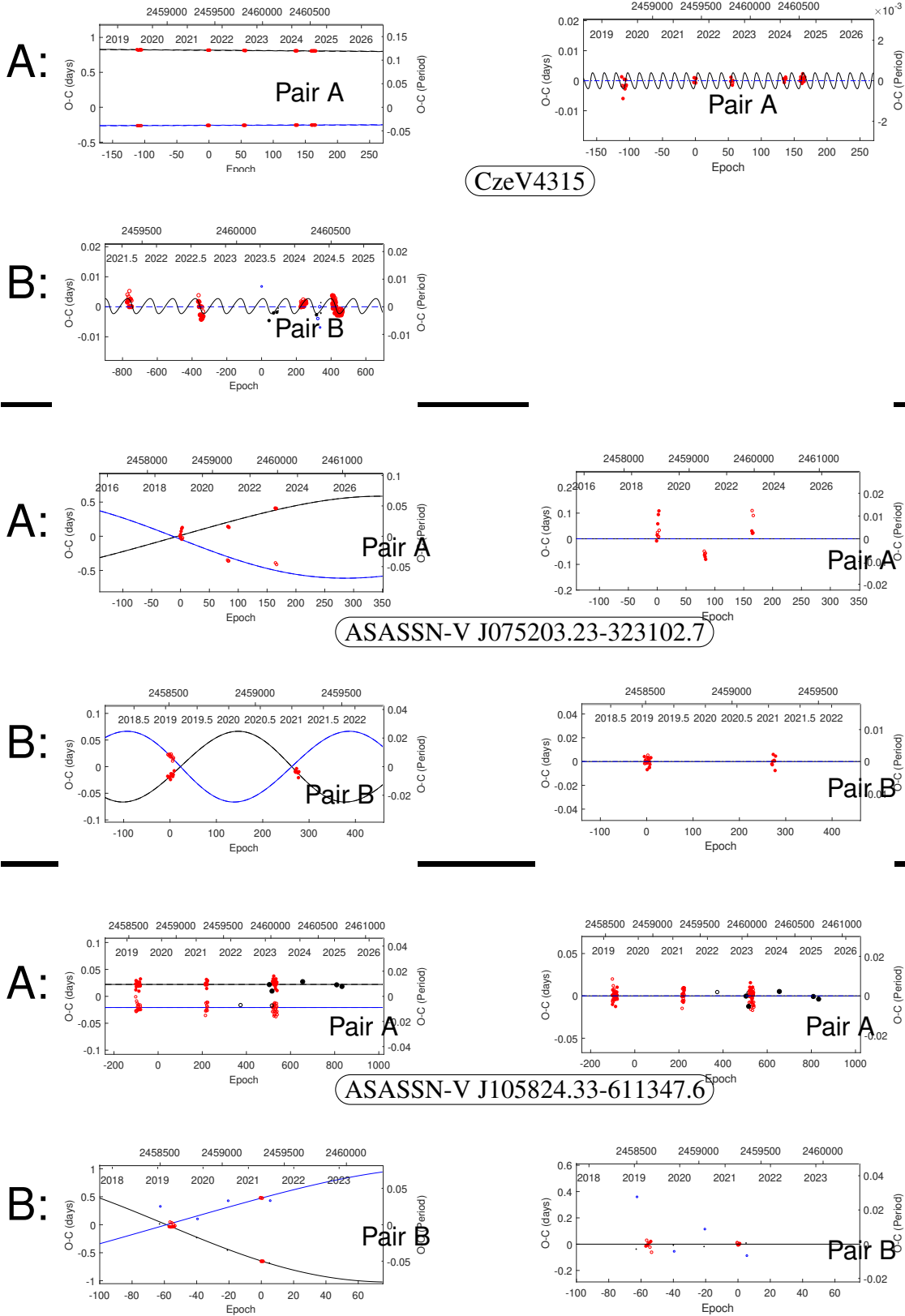

From all the available data, we tried to do the LC analysis of pair B with the special emphasis on the inclination value in particular time epoch. We solved separately all the TESS sectors, and also the individual nights of observations from our ground-based telescopes. These data of resulted inclinations were then fitted with a simple straight line in the (Time-Inclination) diagram and the very approximate value of the rate of inclination change was derived: $\approx$ 1.4$^\circ$/yr
. We are aware of the fact that such an approach is only descriptive ($cos(i)$ should behave like
a sinusoidal variation). But there is still no curvature of these data visible yet, so to fit the
sinusoidal fit to these data would still be very uncertain. All of these derived values of the
inclination angles are given in Table \ref{Tab_CzeV4315_incl}.

Standard theory of triple-star mechanics on the orbital precession caused by the third body was
explained elsewhere in the past (see e.g. \citealt{1975A&A....42..229S}, or
\citealt{2015MNRAS.448..946B}). Only the third bodies orbiting on adequately inclined orbit with
respect to the inner eclipsing binary are able to cause significant changes on the orientations of
the orbits of the inner pair. Another necessary condition is that the outer orbit should be short
enough to produce measurable inclination change.

Having no information about such a body, we can only speculate about its nature. However, there
are in principle two different possibilities. Due to fact that the system is also spatially
resolved (the two stellar images are about 1.9$^{\prime\prime}$ distant), we ask whether we can
rule out the possibility that the two eclipsing signals originate from the same stellar image
(i.e. architecture of the system 2+2 + 1), or are coming each from the different component (i.e.
structure of the system as 2+1 + 2)?

Due to large TESS pixels, this task is not easy. We tried to use the approach introduced in
\cite{2023AJ....165..141H} for localizing the variability sources in crowded fields. However,
without any reliable success (probably due to too large TESS pixels compared to the angular
separation of the two stellar images). Unfortunately, the star was not observed with the ZTF
survey with its higher angular resolution to judge whether both eclipsing signals can be
identified with one or two stellar sources (the star is probably easily too bright for ZTF). Also
our new dedicated observations were not successful at all, having larger scatter due to seeing and
other observational limitations.

Therefore, another possibility would be to detect the ETV of both pairs. If these behave in the
opposite manner, then the system is bound in configuration 2+2 + 1 (this only applies in case of
ETV dominated by pure light-travel-time effect, and necessarily not for systems with strong
dynamical effects). However, to obtain such a data in adequately good quality and over long-enough
time interval is tricky. Especially due to unequal eclipses over time, and also a quite longer
orbital period of the pair A. We used all TESS sectors of data and tried to disentangle the two
eclipsing signals to derive the times of eclipses for both pairs. A similar method was used
recently in our previous publications (see e.g. \citealt{2024A&A...687A...6Z}). For the
shorter-period pair B the detection of a periodic variation in the ETV diagram was quite simple.
See Figure \ref{Figs_ETV} for the result. Also, our new dedicated observations of worse quality
(compared to TESS) indicate such a periodicity. It has a period of about 119 days, which yielded
an order-of-magnitude estimate of a prospective nodal precession (which should have a periodicity
proportional to $p_{AB}^2/P_B$, see e.g. \citealt{2022Galax..10....9B}), i.e. longer than 40
years. However, when we tried to identify a similar behaviour also for pair A, there arises a
problem. At first, a long-term apsidal motion of the pair ($U>3000$ yr) has to be subtracted first
to detect the additional variation. Due to its long period and only very limited number of data
points, one cannot definitely prove its similar periodicity (see Fig.\ref{Figs_ETV}). One has to
take into account that to detect a variation with the amplitude of about a few minutes is much
more tricky for a pair A with its 6.7-d period, than for the pair B with a 0.9-d period. Hence, to
conclude, a confirmation of hypothetical 119-d periodicity from spectra, for example, would be
very fruitful in the future. In Table \ref{LCparamAll} the up-to-date ephemerides are also given
for planning new observations, which would also be very useful for photometry. Therefore, we leave
the true architecture of 2+2 + 1, or 2+1 + 2 as an open question here. A fact what would be taken
as a weak indirect proof favouring slighly the (2+1 + 2) architecture would be the absence of any
rapid apsidal motion of the pair A caused by the strong dynamical interaction.

\subsection{ASASSN-V J075203.23-323102.7}

Another system studied in our sample is ASASSN-V J075203.15-323102.4. This is the star much less
studied, and also a bit fainter than the first one. The star was first detected as a doubly
eclipsing one by \cite{2023MNRAS.520.2386R} who presented its dominant period as 8.8706 days, but
with no value of the second period. That was the reason why the system attracted our attention.
Quite surprisingly, the ASAS-SN catalogue gave its type as gamma Cas, but with an uncertainty flag
and no period (see \citealt{2018MNRAS.477.3145J,2017PASP..129j4502K}). Quite surprisingly the
system also shows a higher RUWE value from the GAIA DR3 catalogue \citep{2023A&A...674A...1G},
namely $RUWE=2.898$, which should also indicate an unresolved binary.

The star, after our inspection of the available TESS data, shows two periodicities. The original
8.87d and also very shallow 2.6817d. Both are of detached type according to their narrow eclipses
well-separated from each other.

\begin{figure}
  \centering
   \includegraphics[width=0.30\textwidth,angle=270]{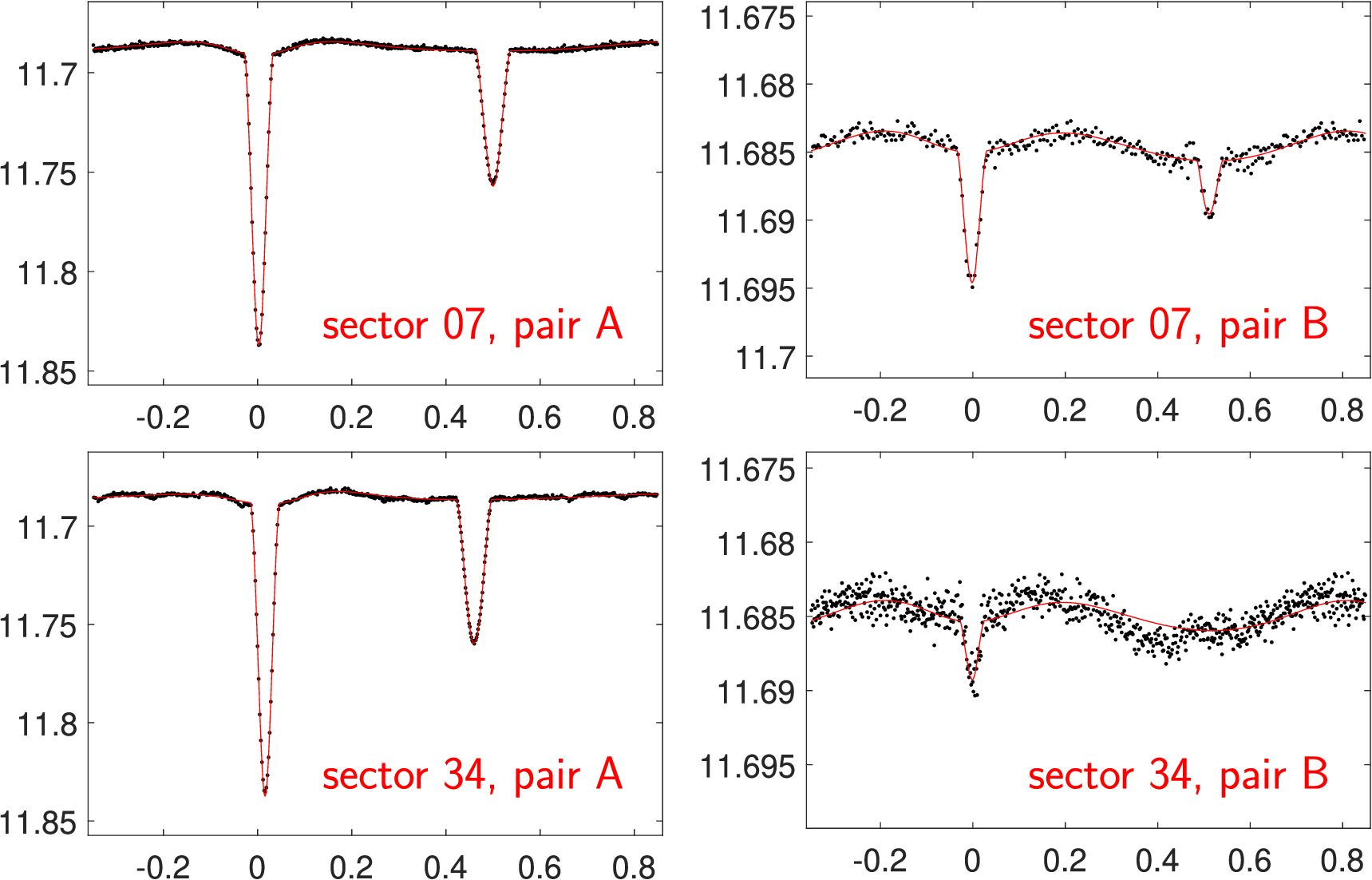}
  \caption{Star ASASSN-V J075203.23-323102.7: Comparison of both light curve shapes of pairs A and B in sectors 07 and 34. The difference of depths of pair B is visible by a naked eye. In sector 61 the pair B completely disappeared.}
  \label{ASAS075203_sectors}
 \end{figure}

However, what is surprising on the TESS data is the fact that the second periodicity is only
visible on sectors 07+08, obviously shallower on sector 34, and later became undetectable. Hence,
the inclination change should be rather rapid and with these available TESS sectors of data the
effect is definitely confirmed. The shape of the individual light curves and their changes in
between sectors 07 and 34 are shown in Fig. \ref{ASAS075203_sectors}. The light curve solution
derived from TESS sector 07 is given in Table \ref{LCparamAll}.

  \begin{figure}
  \centering
   \includegraphics[width=0.4\textwidth]{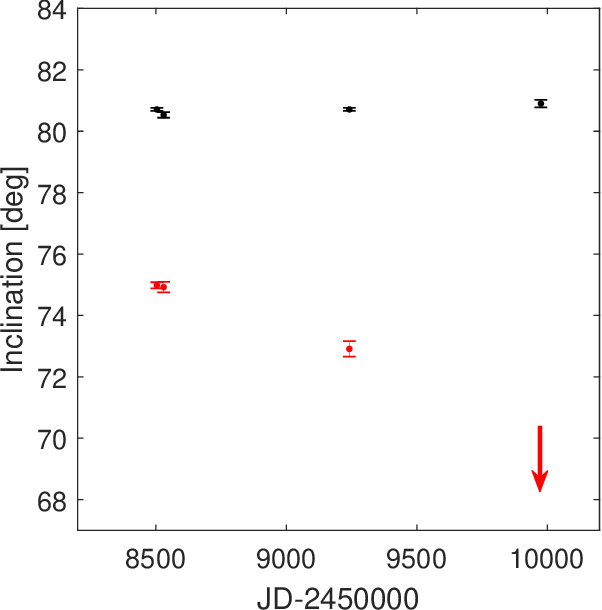}
  \caption{Star ASASSN-V J075203.23-323102.7: inclination angles of both pairs. Black dots for pair A, while red dots for precessing pair B. These data are from 4 different TESS sectors: 07, 08, 34, and 61, respectively.}
  \label{ASAS075203_incl}
 \end{figure}

The rate of inclination change of pair B is fast, see Figure \ref{ASAS075203_incl}. Obviously, the inclination of the pair A remains constant over the whole interval of the TESS data. The last sector 61 gives us only some limitation of the inclination angle for pair B, since the eclipses of B completely vanished. Nevertheless, the rate of inclination advance is so rapid that the linear fit gives us its rough estimate as 1.03$^\circ$/yr. 
From this value a very rough estimation of the suggested A-B period is of about 590 days, which
should be taken as an upper limit for such a period.

As for the previous case, we tried to find an additional proof that our hypothesis is correct.
Detecting the period changes similarly as for CzeV4315 is here more complicated. We deal here with
a very limited number of usable data points, very shallow eclipses of pair B, and also a long
orbital period of pair A. All of this makes any reliable analysis still rather problematic. The
pair A (the dominant one with 8.87-d period) shows in the ETV diagram also large variations as
well as fast apsidal motion (it has the eccentricity of about 0.21, see Fig.\ref{Figs_ETV}). Fast
ETVs are also visible during one TESS sector of data, i.e. period change should be rapid. This
should indicate large dynamical interaction between the bodies and should serve as an indirect
evidence for possible inclination changes. From their amplitude of the order of 0.1 days one is
able (due to relation of amplitudes of dynamical ETV $\Delta_{dyn} \sim P_{A}^2/P_{AB}$) to
roughly estimate the potential mutual orbit of the order of 500-1000 days (i.e. in rough agreement
with the $P_{AB}$ estimate above). However, the pair B has only limited data from TESS, moreover
in sector 34 only for the primary eclipses. With such a dataset finding a reliable solution is
still problematic. However, according to our modelling the apsidal motion is very rapid (order of
magnitude period of apsidal advance only about 3.5yr). Nevertheless, such a rapid apsidal advance
cannot be attributed to the classical mechanics and the tidal interaction between the components,
neither to the relativistic contribution (both these effects together should led to apsidal motion
of the order of magnitude slower). Hence, this can also be taken as an indirect evidence of the
dynamical interaction. Linear ephemerides for both A and B binaries are also given in Table
\ref{LCparamAll}, which would be useful for planning new observations (still important to confirm
our hypothesis in the following years).

\subsection{ASASSN-V J105824.33-611347.6}

The third system in our sample is named ASASSN-V J105824.33-611347.6 ( = TIC 465899856). It will
be only very briefly mentioned here, since we still have only very limited information about the
star. Despite the target is a bit fainter than the previous ones, it was classified
spectroscopically and its spectral type is the earliest in our sample, being about O6.5IV
\citep{2022A&A...657A.131M}. Close to the target itself, there was found a star named UCAC4
144-063899 = TIC 465899824 (about 12$^{\prime\prime}$ distant). This star is slightly fainter than
ASASSN-V J105824.33-611347.6, but share a comparable parallax and proper motion as well. Hence, we
speculate that these are probably weakly gravitationally bound.

\begin{figure}
  \centering
   \includegraphics[width=0.45\textwidth]{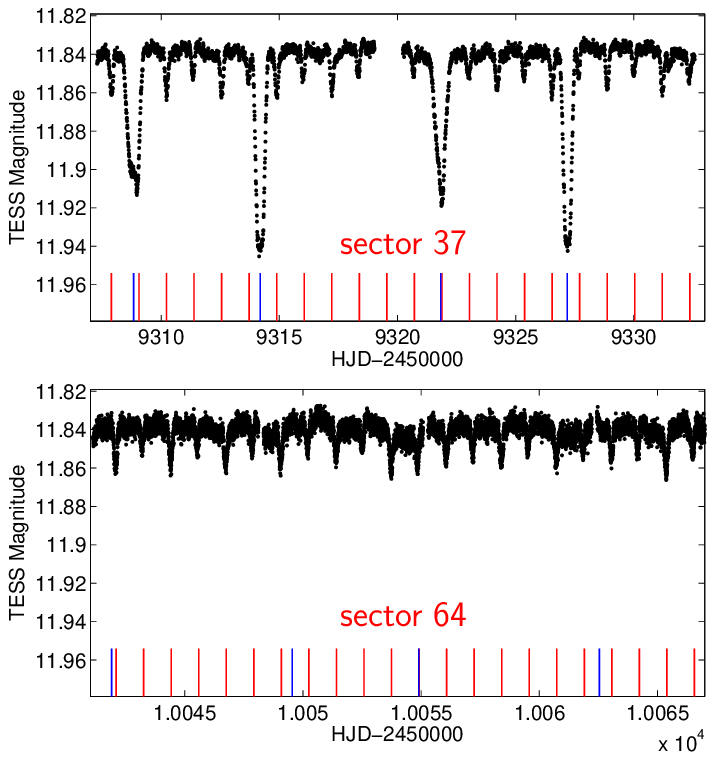}
  \caption{Star ASASSN-V J105824.33-611347.6 and its TESS photometry from two different sectors. As one can see, the dominant 13-day eclipses disappeared completely. The lower abscissae indicate the times of mid-eclipses for the pair A (2.33-d period, in red), and for the pair B (13-d, blue).}
  \label{ASAS105824}
 \end{figure}

Quite surprisingly, the star was not included into the OGLE database of variable stars despite its
coordinates are within the limits for the OGLE survey in galactic disc. On the TESS data there
appears two periodicities. On sectors 10, 11, and 37 there are the dominant eclipsing pair with
its 0.1 magnitude deep eclipses and period of about 13 days (we call this pair B), and also the
shallower eclipses  A of about 0.02 mag only and period of about 2.33 days. Both inner eclipsing
pairs are obviously well-detached systems. All of this can be seen in our Figure \ref{ASAS105824}.
There also can be seen that the most recent sectors (63 and 64) do not show any 13-day eclipses,
which easily disappeared. The inclination change should be rather fast.

\begin{figure}
  \centering
   \includegraphics[width=0.45\textwidth]{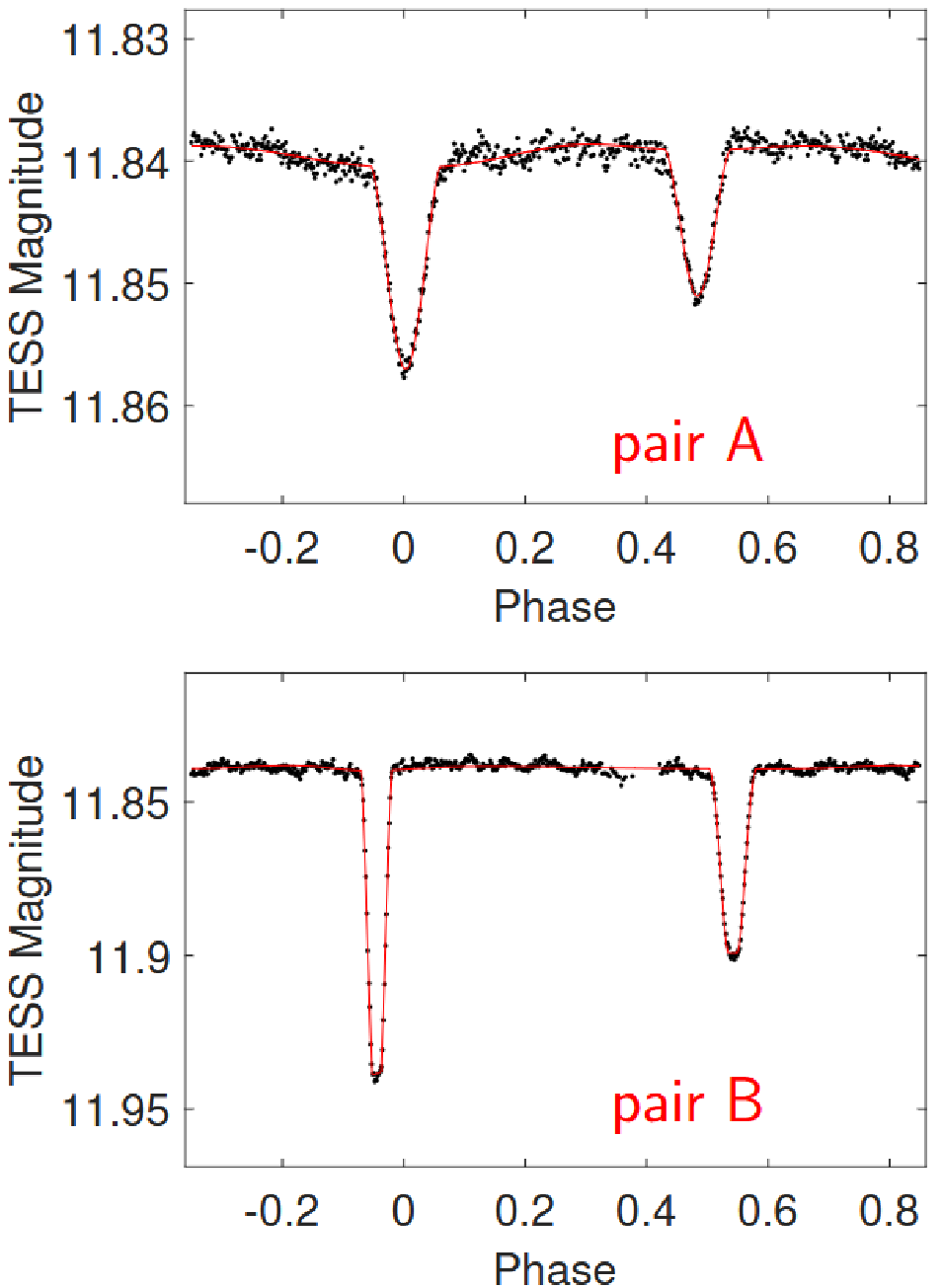}
  \caption{Both A and B pairs of ASASSN-V J105824.33-611347.6 as derived from the TESS photometry (sector 37). The light curve fitting parameters are given in Table \ref{LCparamAll}.}
  \label{ASAS105824LCs}
 \end{figure}

Our preliminary light curve fitting was done on TESS sector 37 with the most pronounced eclipses
of both pairs. See the Table \ref{LCparamAll} for the parameters and also Figure
\ref{ASAS105824LCs} for the fits. Both inner pairs show eccentric orbits.

\begin{figure}
  \centering
   \includegraphics[width=0.45\textwidth]{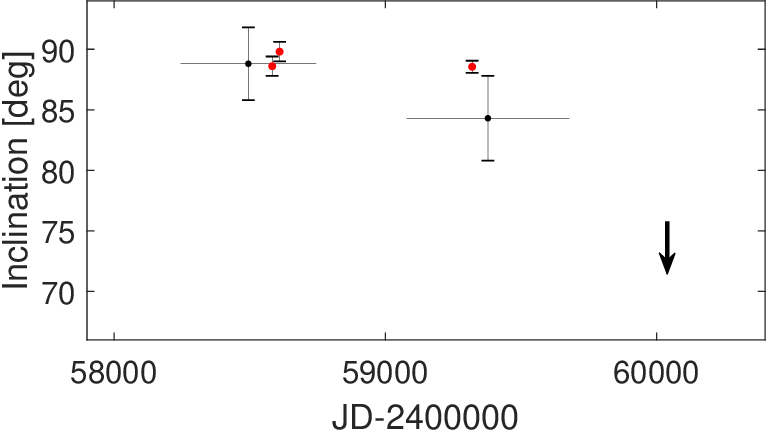}
  \caption{Inclination change of ASASSN-V J105824.33-611347.6 as derived from TESS sectors (red dots), together with the estimations from the ASAS-SN data (black dots).}
  \label{ASAS105824incl}
 \end{figure}

What is also quite remarkable is a fact that in older ASAS-SN photometry there can be seen the
longer pair with 13-d period, but only very hardly the shorter 2.33-d variation. On the other
hand, on even older photometry from ASAS-3 there cannot be seen any of these two variations. A
natural explanation of these non-detections of eclipses is easily the fact that the eclipses are
rather shallow, especially the shorter 2.33-d one. In our Figure \ref{ASAS105824incl} we plot the
inclination changes of the pair B as derived from the individual TESS sectors of data, and besides
that also some rough estimation of the inclination as resulted from the ASAS-SN photometry.
However, it is of poor quality only and barely usable for any analysis.

The same as for the previous system also applies here: new observations are needed. Photometry, as
well as spectroscopy can tell us much more about this interesting system. The 13-day subsystem is
expected to be the dominant one, so its components should be detected in the spectra despite the
fact that nowadays this pair is non-eclipsing.

We tried to follow a similar approach as for the previous cases here. However, having only limited
data in our hands, one can only hardly do any reliable analysis. The results are given in our
Fig.\ref{Figs_ETV}, where one can see no obvious ETV for pair A (apart of the long term apsidal
motion causing only shift between primary and secondary eclipses). Also our new observed data from
the Danish 1.54-m telescope are not very usable in this aspect (i.e. the black symbols in
Fig.\ref{Figs_ETV}). However, on the other hand, the pair B shows rapid ETV variations, which
cannot be easily attributed to the classical apsidal motion and are probably caused by some
quadruple-star dynamics. Hence, we can take this as an indirect proof of our hypothesis.

\section{Conclusions}  \label{Conclusions}


The quadruple stellar systems of 2+2 architecture are very important for our knowledge about the
multiple star formation theories (see e.g. \citealt{2021Univ....7..352T}). Moreover, if these host
eclipsing binaries within both inner pairs, our set of derivable parameters of all components
increases, and we can compute most of the orbital and physical parameters in such complicated
systems. However, what is usually missing is some information about the mutual orbit of the two
pairs around a common barycenter, and the mutual inclination angles between the individual orbits.
In our work, we present a unique group of 2+2 doubly eclipsing quadruples, where one of the inner
eclipsing binaries is changing its orientation in space. Thanks to such unique systems, we can
model its dynamics \citep{2022AJ....163...94V} fully, even judging which of the
prograde/retrograde orbital motion is the correct one (if we have enough data in the future).

We are not aware of any such systems on the whole sky nowadays. However, one should expect that
thanks to ultra-precise data available from space, like the recent TESS observations, there will
be discovered other similar examples like these in the future. What should also be mentioned is an
independent confirmation of our hypothesis by other methods. Namely the spectroscopy would be very
useful (for precise absolute parameters of the components, as well as radial velocity variations
on the long mutual orbit), but also the interferometric detection of the double would bring us
additional information. Unfortunately, all the systems are relatively distant (all of them
$>$1kpc), and their orbital periods are rather short (hundreds of days estimated), therefore the
most suitable is the first one system CzeV4315, which still has the predicted angular separation
of only 1.5mas. To conclude, lets take this our study as a call for the observers to focus their
attention to these interesting systems, especially the first one CzeV4315, which is getting closer
to 90$^\circ$ hence its eclipses are getting deeper and deeper in the next few years.

For all of our three systems there resulted that the level of the third light from the two
separate light curve solutions is well above the expected 100\%. However, this is not such a
surprising fact, since we used the TESS photometry with a poor angular resolution for one pixel.
Moreover we also used typically 3x3 pixels for extracting the photometry, hence rather large area
around a target, where also some close-by sources usually contaminate our target star as well.
Therefore, the contribution of additional light can be identified with these close stars.

For the last one from our studied systems, the result is that the pair showing the inclination
changes is the one with the longer orbital period, in contrast with the two previous cases. One
can ask, whether it is even possible. Yes, it is, since the effect of orbital precession depends
on the overall architecture of the system. Not only the individual masses, but also mainly the
three individual angular momenta of the three orbits, and their orientations play a crucial role
here. If the orbits are not co-planar, then the ratio of the momenta (i.e. the mutual orientations
of the orbits with respect to each other) is the key factor for the precession. Therefore, also
the architecture with the precessing longer pair is possible. Unfortunately, now we do not have
any information about the mutual inclinations of the orbits, so we cannot say anything about that.
One possible solution would be to resolve the pair interferometrically, to derive the architecture
of the whole quadruple.

\section*{Acknowledgments}
We do thank an anonymous referee for his/her helpful and critical suggestions, which significantly
increased the level of the whole manuscript and its findings. We do thank the ASAS, ASAS-SN, OGLE,
and TESS teams for making all of the observations easily public available. We are also grateful to
the ESO team at the La Silla Observatory for their help in maintaining and operating the Danish
1.54m telescope. The research of P.Z., J.K., and J.M. was also supported by the project {\sc
Cooperatio - Physics} of Charles University in Prague. The research of JM was supported by the
Czech Science Foundation (GACR) project no. 24-10608O. MM would like to thank to projects financed
by Ministry of Education of the Czech Republic LM2023032 and LM2023047. The observations by ZH in
Velt\v{e}\v{z}e were obtained with a CCD camera kindly borrowed by the Variable Star and Exoplanet
Section of the Czech Astronomical Society. This research made use of Lightkurve, a Python package
for TESS data analysis \citep{2018ascl.soft12013L}. This research has made use of the SIMBAD and
VIZIER databases, operated at CDS, Strasbourg, France and of NASA Astrophysics Data System
Bibliographic Services. This work has made use of data from the European Space Agency (ESA)
mission {\it Gaia} (\url{https://www.cosmos.esa.int/gaia}), processed by the {\it Gaia} Data
Processing and Analysis Consortium (DPAC,
\url{https://www.cosmos.esa.int/web/gaia/dpac/consortium}). Funding for the DPAC has been provided
by national institutions, in particular the institutions participating in the {\it Gaia}
Multilateral Agreement.

\section*{Data availability}

All the data used in the manuscript, which are not already included in the Tables, will be shared
on reasonable request to the corresponding author.

\end{document}